\documentclass[reprint, amsmath,amssymb, aps, prl]{revtex4-2}
\usepackage[utf8]{inputenc}
\usepackage{graphicx}
\usepackage{bm}
\usepackage{hyperref}
\usepackage{xcolor}
\usepackage[normalem]{ulem}
\usepackage[ruled,lined]{algorithm2e}
\usepackage{setspace}
\usepackage{xcolor}

\definecolor{crimson}{HTML}{DC143C}
\definecolor{perfect_green}{HTML}{4FBF26}

\def\equationautorefname~#1\null{Eq.~(#1)\null}
\def\figureautorefname~#1\null{FIG.~#1\null}

\begin{document}

\title{Critical demand in a stochastic model \\of flows in supply networks}

\author{Yannick Feld$^1$}
\author{Marc Barthelemy$^{1,2}$}
\affiliation{$^1$Université Paris-Saclay, CNRS, CEA, Institut de
  Physique Théorique, 91191 Gif-sur-Yvette, France}
\affiliation{$^2$Centre d'Analyse et de Math\'ematique Sociales (CNRS/EHESS) 54 Avenue de Raspail, 75006 Paris, France}

\date{\today}

\keywords{Supply chains, criticality, material flow, network topology}

\date{\today}

\begin{abstract}

Supply networks are essential for modern production, yet their critical properties remain understudied. We present a stochastic model with random production capacities to analyze material flow to a root node, focusing on topology and buffer stocks. The critical demand, where unsatisfied demand diverges, is examined mostly through numerical simulations. Without stocks, minimal production dictates behavior, making topology irrelevant. With stocks, memory effects arise, making topology crucial. Increased local connectivity is beneficial: firms should favor broad, short supply chains over long, narrow ones.

\end{abstract}

\maketitle


Supply networks are crucial for modern production and, by extension,
for society as a whole \cite{nagurney2013, nagurney2023, nagurney2024}. Firms rely on timely
input delivery—such as a screw factory needing metal—to sustain production. Inputs may be
stored in buffer stocks or delivered just-in-time \cite{Bartezzaghi2016,
  Katsaliaki2022}. Buffer stocks improve resilience but add costs, creating a trade-off between efficiency and
disruption resistance, a critical factor in supply network
optimization. Supply chains are often modeled using flow equations
\cite{nagurney2013, nagurney2023}, where nodes represent firms or warehouses and links represent product
flows. Typically, models solve for optimal cost efficiency, often via variational
inequalities \cite{dong2004, nagurney2013, nagurney2023}. While effective for
management, these models lack insights into disruption responses, which can range
from minor delays \cite{Colon2017} to major crises
\cite{Caraiani2024}. A key question is whether the system
can absorb disruptions or if they trigger cascading failures
\cite{battiston2007, Hearnshaw2013}.

Complex, real-world models \cite{Ramanathan2020, sawik2024} often
obscure underlying resilience principles, and Moran et
al.~\cite{moran2024} introduced recently a
stochastic model to study delay propagation in supply networks with dynamic topology,
revealing critical behavior where delays escalate beyond a
threshold. However, this model does not include key factors such as demand and stocks.
Here, we present a stochastic model of material flow in a supply network,
focusing on a single \emph{root} firm selling to consumers. Our model examines how
stochastic supply fluctuations impact material flow, particularly how buffer
stocks influence system dynamics. We will focus on the critical demand rate $r^*$,
above which the system transitions to a regime where it is no 
longer able to meet the demand of customers. Stocks add a memory effect, known in statistical
physics to alter system dynamics in general, and we expect them to strongly affect the critical
demand rate $r^*$. By studying how
network structure impacts this critical rate, we aim to identify design principles for resilient supply networks.\\

\begin{figure}[h!]
    \includegraphics[width=\linewidth]{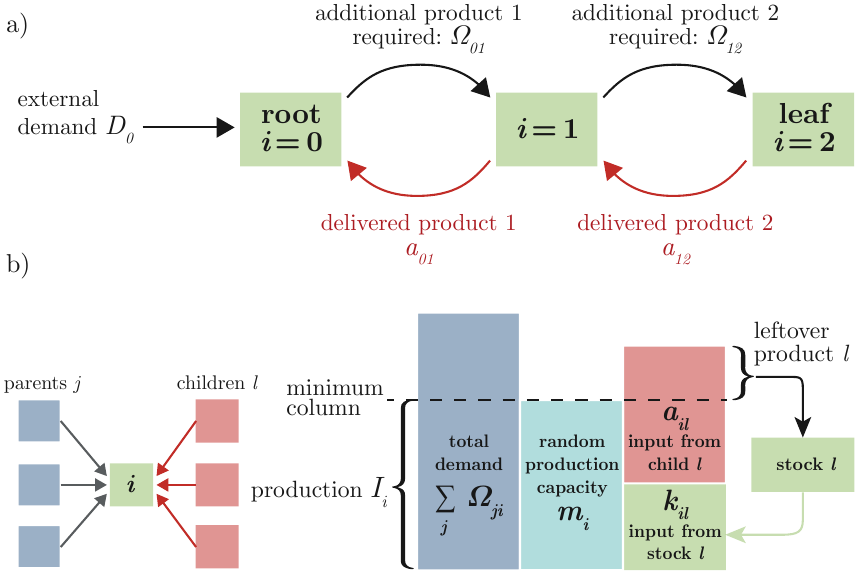}
    \caption{ \label{schematic} Schematic depiction of the dynamics. In a) we see a 
    small example network with a root node $i=0$, an intermediate node $i=1$ and 
    a leaf node $i=2$. The flow of demand and product is visualized.
    In b) we show a zoom on an arbitrary node $i$ with parents $j$ and children $l$. 
    It visualizes the production $I_i$ as a result of the demand, the random 
    production capacity $m_i$, the delivered product $a_{il}$ and the stock $k_{il}$.
    }
  \end{figure}

\textit{The model---} We make the fundamental assumption that the material flow network can be represented as a directed acyclic network \cite{Meixell2005, Wiedmer2021, Kito2014, Perera2018}. This implies that no node can be its own ancestor (see Supp.~Mat. for a more detailed explanation). Notably, this assumption excludes processes such as recycling or refurbishing \cite{Fleischmann2000, Francas2008}. The network topology is
represented by a list of $N$ nodes (firms), and each node $i$ has a
list  $P_i$ of parents (the direct customers), and a list $C_i$ of children
(the direct suppliers). We are focusing on one specific product,
intended for external customers,
which is produced by exactly one \emph{root} node 
(referred to as node $i=0$),  which is the common ancestor of all
other nodes. This root node is subject to an external demand $D_0$,
which represents the amount of product ordered by external sources.
The demand $D_0(t)$ at time step $t$ consists of the unsatisfied demand $u(t-1)$, from the previous
time step $t-1$ and the new demand $r$ that arises at the current time
step
\begin{equation}
    \label{root_demand_update}
    D_0(t) = u(t-1) + r~.
  \end{equation}
  (we assume that $r$ is constant). We denote by $D_i$ the demand of an arbitrary firm $i$ which means
that $D_i$ units of product are needed from all its children. Note that the amounts are expressed in arbitrary units, meaning that a demand of 1 could, for example, correspond to requesting 3 tonnes of product from child A but only 10 grams from child B. Each
firm $i$ is able to maintain a stock $k_{il}(t)$ of its child $l$, which is limited by the maximal 
storage capacity $s$, assumed to be the same for all firms. If the stock $k_{il}(t)$ of the product from child $l$
is less then the demand $D_i(t)$, the node $i$ will need to order the remaining amount
\begin{equation}
    \Omega_{il}(t) = \max \left(0, D_i(t) - k_{il}(t)\right)~.
\end{equation}
This mechanism propagates the demand throughout the network, and the
demand of all nodes $i>0$ is
given by summing up the demand they receive from their parents
\begin{equation}
    \label{demand_update}
    D_i(t) = \sum_{j \in P_i} \Omega_{ji}(t)~.
\end{equation}
Due to physical limitations such as the amount of production lines,
employee illness, machinery maintenance, or other operational
requirements, firms have a maximal \emph{production capacity} 
$m_i(t)$ which corresponds then to 
the maximal amount of product that node $i$ can produce at time
$t$. We will assume that these quantities $m_i(t)$ are
independently and identically distributed according to a
uniform distribution over the interval $[0,1]$ (the maximum production
capacity thus sets the scale of the other quantities). These
production capacities are redrawn from the
same distribution each time step (annealed). Leaf nodes do not depend on any other nodes and their
production $I_i(t)$ is equal to $\min[m_i(t), D_i(t)]$, ensuring the nodes will not produce 
more than was ordered. For all other nodes $i$, the production is additionally constrained by the availability of product from their children $l$, 
which consist of the stock $k_{il}(t)$ and the current delivery
$a_{il}(t)$ from $l$ to $i$.
The resulting equation for the production of node $i$ is then given by 
\begin{align}
    \label{production_eq}
    I_i(t) = \min_{l \in C_i}\left[m_i(t), D_i(t), a_{il}(t) + k_{il}(t)\right]~.
\end{align}
The amount of product that a firm can supply to its parents, i.e., customers, at time $t$ is clearly limited by the production $I_i(t)$.
If a firm has multiple parents, the firm has to decide how to split
its production between them, and we assume that the product delivered by node $l$ to its parent $i$ is proportional 
to the amount that node $i$ contributes to the demand $D_l(t)$
\begin{align}
    \label{eq_aij}
    a_{il}(t) &= \begin{cases} 0 &\text{if~} D_l(t)=0 \text{~or~}l\notin C_i \\\frac{I_l(t) \Omega_{il}(t)}{D_l(t)} &\text{otherwise}\end{cases}
\end{align}
Next, each node $i$ will attempt to store any leftover product $l$ for future use by adding it to its stock $k_{il}$. 
Since the maximal storage capacity for each product is $s$,
this leads to the equation
\begin{align}
    \label{eq_kij}
    k_{il}(t+1) &= \begin{cases} \min \left[s, a_{il}(t) + k_{il}(t) - I_i(t)\right] &\text{if~}l\in C_i \\ 0 &\text{otherwise} \end{cases}
\end{align}
The unsatisfied demand of the root node is then $u(t) = D_0(t) -I_0(t)$. 
Since the network is acyclic we can order the above equations for the
different nodes in a way that makes them calculable. In the
Supp. Mat. we provide the pseudo-code for this model and in Fig.~\ref{schematic}, we show the most important interactions of this model.
The main question for this model is to investigate its critical
behavior and to find the critical root demand rate $r=r^*$, beyond which 
the unsatisfied demand $u$ grows on average (see the Supp. Mat. for how
to measure  $r^*$). This average growth
signals that the supply network is unable to keep up with the external
demand. In particular we are interested in how this critical $r^*$ depends on the network topology.\\

\textit{No stocks---}We first investigate this model in the limit of
no stocks obtained for $s=0$,
when all stocks $k_{ij}(t)$ are always 0. We denote by $ {\cal C}_i$ the set of all descendants of node $i$,
including all nodes on which node $i$’s production depends, either directly or indirectly. We focus on the case where all nodes have one parent
(except for the root node), which means that all the demands $D_i$ are
equal, and that $a_{il}(t) = I_i(t)$ if $l$ is the child of $i$. The
equation \autoref{production_eq} for the production becomes
\begin{align}
  \nonumber
  I_i(t)& = \min_{l \in C_i}\left[m_i(t), D_i(t), a_{il}(t)\right]\\
  &= \min_{w \in {\cal C}_i}\left[m_i(t), m_w(t), D_i(t)\right]~.
  \label{line_production_eq}
\end{align}

For a simple one-dimensional structure, each node $i$ (except for the leaf node), has exactly one child $l=i+1$. If the demand is large, the term $D_i$ can be ignored in the minimum in \autoref{production_eq} (and therefore also in \autoref{line_production_eq}). The average production of node $i$ reduces then to the average of the minimum of $N-i$ independent and
identically distributed (i.i.d.) uniform random variables
from the interval [0,1]. The result is known from statistics and is given by
\begin{equation}
    \label{no_stock_i_eq}
    \left<I_i\right> =\frac{1}{1+N-i} ~.
\end{equation}
Interestingly, this means that the production only depends on the distance to the leaf.
Clearly, the critical root demand $r^*$ is given by the average
production of the root node $i=0$, which implies that
\begin{equation}
    \label{No_stock_crit_eq}
    r^*=\frac{1}{N+1}~.
  \end{equation}
Note that the latter equation is true for any network where each node has at most one parent, i.e., 
for all tree-like networks (see the discussion in the
Supp. Mat.). Interestingly, for the 1D chain in the quenched case for
the $m_i$'s, the root node demand evolution is given by the stochastic equation
\begin{align}
  \frac{dD_0}{dt} = -\min(\{m_i\}, D_0) + r,
\end{align}
highlighting the production bottleneck defined by the smallest $m_i$.

\begin{figure}[h]
    \centering
    \includegraphics[width=\linewidth]{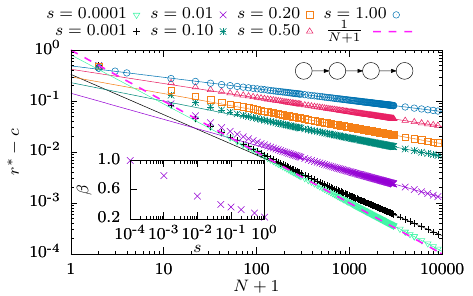} 
    \caption{
        \label{different_s_fit} Numerical results for the critical root
        demand $r^*$ shifted by $c$ versus the number of nodes $N$ for a single chain
        for different stock capacities $s$. 
        The symbols represent the measured data, while the lines represent the fits of \autoref{eq_fit_s_single_chain},
        whereas the dashed line 
        represents the case without stock
        \autoref{No_stock_crit_eq}. The small graph displayed in the
        plot
        is a visual indication that this plot is about the single chain.
        The inset shows $\beta$ of the fits versus $s$. All coefficients of determination $R^2$
        are larger than 0.998, indicating a good fit quality. 
    }
\end{figure}

\textit{Non-zero stocks---}We now examine non-zero stock capacities ($s > 0$), which give the
system some memory, complicating analytical treatment.
In the 1D chain, some analytical results are available (see
Supp. Mat.),
but we primarily study the model through numerical simulations \cite{hartmann2015}. To explore the impact of stock capacity $s$, we measure the critical
root demand $r^*$ versus the number of
nodes $N$ (at a fixed value of $s$). Notably, node $i$'s ancestors do not affect its dynamics,
so in a simple chain, average production
depends only on distance to the leaf. For large $N$, data fits a power
law of the form
\begin{equation}
    \label{eq_fit_s_single_chain}
    r^* = \frac{\alpha}{(N+1)^\beta} + c,
\end{equation}
with $\alpha$, $\beta$, and $c$ as fitting parameters depending on
$s$. Results are presented in Fig.~\ref{different_s_fit}, where data for $N \geq 400 $ is used in the fits to reduce finite-size effects.

As with no stock, increasing $N$ decreases $r^*$, as more dependencies
hinder production. For $N = 1$, stock
is irrelevant, and $r^* = \frac{1}{2}$. For larger $N$, increasing $s$
raises $r^*$ since fluctuations are buffered
by stocks $k_{il}$. We emphasize that stocks not only modify the prefactor but also alter the exponent, fundamentally changing the system's behavior. This effect is particularly significant for long supply chains, where incorporating stocks provides a major advantage. For $s \to 0$, we recover \autoref{No_stock_crit_eq} with $\beta \to
1$ (see inset of Fig.~\ref{different_s_fit}), $\alpha \to 1$, and $c
\to 0$ (see Supp. Mat. Fig.~S9).

For large $s$, we expect $r^* \to \frac{1}{2}$, as surplus production
is stored, leading to an average node
production of $\frac{1}{2}$. However, with empty initial stocks,
stable production at $\frac{1}{2}$ may take
time, depending on chain length and numerical results for very large $s$ align with
this (Supp. Mat.). This behavior changes if any node has multiple
parents,
where production splits between them. In the rest of the paper, we
will focus on ere and in the fo, we focus on $s=1$.\\

\textit{Open and closed chain structure---}We now investigate the effect of an open chain structure, where $z_0$
chains of the same length start from the root node. 
We define the height of a
node $j$ here as the number of nodes in the shortest path from $j$ to the root
node.  Thus, the height of the root node is $h=1$.
The height of the network is defined as the maximum height of all nodes.

We measure the critical root demand $r^*$ for various $z_0$ and
different heights $h$, and use the same fitting function
\autoref{eq_fit_s_single_chain} (see the Supp. Mat.). We basically
find that networks with more 
chains converge to a lower critical demand rate $r^*$ for $N \to \infty$, which makes intuitive 
sense as there are more dependencies throughout the network.

For the closed chain case, we assume that all these chains end up at
another node $\theta$, and we also add a new chain to $\theta$,
allowing us to vary the distance of the connecting node $\theta$ to
the leaf. The most important topological feature of this structure is that 
multiple nodes have the common supplier $\theta$.
The product of this supplier has to be divided between $z_0$ nodes, which means that the critical root demand is limited by
$r^* \leq \frac{\left<I_\theta\right>}{z_0}$
i.e., the network is unable to produce more than what node $\theta$ is able to produce divided by the number of its parents.\\

\begin{figure}[htb]
    \centering
    \includegraphics[width=\linewidth]{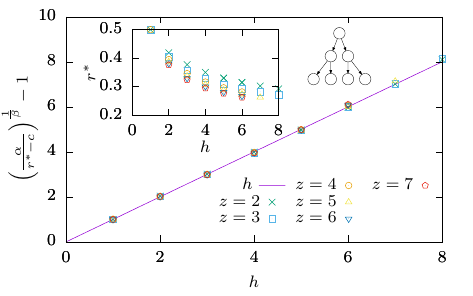} 
    \caption{\label{tree_collapse_fig} We show here the data collapse obtained by fitting \autoref{tree_collapse_eq} to the numerically measured critical
    root demand $r^*$ for different regular trees defined by their child counts $z$ and heights $h$. The coefficients of determination $R^2$ were all larger than 0.999. The little network in the plot is a visual aid to indicate that this plot is about tree structures. The inset shows $r^*$ over $h$ for the sake of completeness.
    }
\end{figure}

\textit{Regular tree---}We now focus on a regular tree structure, where each node branches into $z$ children until 
the height $h$ of the tree is reached. We observe numerically
that the critical demand is well described (in the range of value of $h$ that are accessible numerically) by the following function
\begin{align}
    \label{tree_collapse_eq}
    r^*(z, h) = \frac{\alpha}{\left(h+1\right)^{\beta}}+c~,
\end{align}
where $\alpha$, $\beta$ and $c$ are fitting parameters depending on $z$. As required, Eq.~\ref{tree_collapse_eq} reduces to Eq.~\ref{eq_fit_s_single_chain} for $z=1$. In order to test this function, we note that it implies that
plotting $\left(\frac{\alpha}{r^*-c}\right)^\frac{1}{\beta} - 1$ versus $h$
should be the straight line $y=x$ for all values of $h$ and $z$. We then obtain the plot shown in
Fig.~\ref{tree_collapse_fig} displaying an excellent data collapse,
confirming the relevance of the fitting function (fitting parameters
can be found in the Supp. Mat.~Fig.~S10). 
We observe here that for increasing $z$, the exponent $\beta$
increases also. Indeed, a larger $z$ means more nodes, therefore more
dependencies and a smaller critical rate.
In contrast $c$ seems to stay relatively constant, which implies that 
for $N \to \infty$ all of the measured trees converge to the same 
critical demand rate. The critical rate $r^*$ appears to be much
larger than the single chain case for $s=1$, but this result is
difficult to confirm numerically as the number of nodes increases
exponentially with $z$. \\

\begin{figure}[htb]
    \centering
    \includegraphics[width=\linewidth]{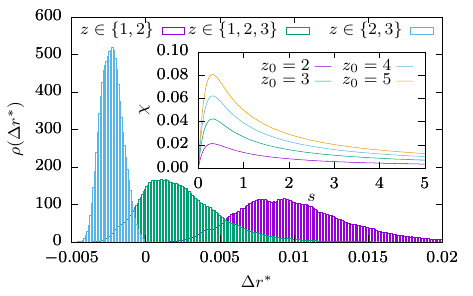}
    \caption{ \label{rand_vs_reg_tree_plot} On the main plot, we show the numerically obtained probability density 
    $\rho(\Delta r^*)$ for different sets for the distribution of $z$
    (with a maximal stock capacity of $s=1$). For each probability density we drew $8 \times 10^5$ independent random trees and measured their
    respective critical root demand $r^*$ to create the histogram. The violet bar plot corresponds to $z \in \{1,2\}$, the green one to $z \in \{1,2,3\}$ and the   
    blue one to $z \in \{2,3\}$. All trees used here have exactly $N=63$ nodes corresponding to a height of $h = 6$.
    The inset shows the measurement of the production difference $\chi$ versus the stock $s$ for different minimal trees, which consist of the root node 
    and its $z_0$ children, i.e., $N=z_0+1$.
    }
\end{figure}

\textit{Random tree structure---}We now examine the impact of randomness on a tree structure by
comparing a regular tree and a random tree, both containing the same
number of nodes. To generate random trees with $N$ nodes, we begin
with the root node (layer 1). For each node in the current layer, the
number of children is randomly selected from a predefined set, such as
$\{1, 2, 3\}$, and children are added one by one. This process
is repeated layer by layer, with new children assigned until adding
more nodes would exceed the target of $N$. This ensures the tree has
exactly $N$ nodes. The effect of randomness is quantified by
comparing the critical root demands of the random and the regular tree
\begin{equation}
    \Delta r^* = r^*_{regular} - r^*_{random}
\end{equation}
A positive \(\Delta r^*\) indicates that the regular tree has a
higher average production capacity, whereas a negative \(\Delta r^*\)
suggests that the random tree performs better. The distribution of
\(\Delta r^*\) for different cases are shown
in Fig.~\ref{rand_vs_reg_tree_plot}.
We observe that random trees with local connectivity $z \in \{1,2\}$
are performing worse than the regular one, while those created with
$z \in \{2,3\}$ are always performing better 
than the regular tree. Random trees with $z \in \{1,2,3\}$ typically
have performance in between these two cases. These results show that in general
it is beneficial to avoid linear segments and to keep
descendants as close to the root as possible.

To investigate further, we compare a tree structure with a root node
and $N-1$ children against a line of $N$ nodes,
with equal production capacities in both cases. We use a root demand
greater than 1, ensuring the system
is not demand-limited (details in Supp. Mat.). We measure the production gain of the tree over the line as
\begin{equation}
    \chi = \frac{I^{tree}_{tot} - I^{line}_{tot}}{I^{line}_{tot}}~.
\end{equation}
Results in the inset of Fig.~\ref{rand_vs_reg_tree_plot} show that for
$s=0$, $\chi=0$, as expected since
only the number of nodes matters without stocks.
For all values of $s$, $\chi$ is positive
and is growing with $z_0$ demonstrating the advantage of a tree-like
structure over a line.  As $s$ increases, $\chi$
peaks at $s \approx 0.315$. For large $s$, $\chi$ returns to $0$ because stocks are effectively
unlimited, preventing waste. Thus, for tree-like topologies, the average production approaches the
capacity, i.e., $\frac{1}{2}$,
since demand is not limiting.\\

\textit{Concluding remarks---}In this paper, we introduced a stochastic model for material flow in supply chain networks and investigated the critical demand rate $r^*$. When stocks are absent ($s=0$), our analytical derivation shows that the critical demand rate $r^*$ is trivial and independent of the network topology, provided each node has at most one parent.

With the inclusion of stocks ($s > 0$), a memory effect emerges, leading to more complex dynamics where network topology becomes relevant. Except for a special case detailed in the Supplementary Material, this scenario becomes too intricate for analytical treatment and is primarily studied numerically. Stocks fundamentally alter system's behavior, affecting not only the prefactor but also the exponent of the critical demand rate. For real-world supply chains, even small stockpiles within the network provide significant benefits for long supply chains and should be prioritized over just-in-time delivery.

We examined a variety of relevant topologies, ranging from simple chains to multiple chains, as well as regular and random trees. Our simulations indicate that, in general, keeping nodes closer to the root is beneficial, favoring tree-like structures over extended linear segments. As a practical guideline, firms restructuring their supply chains should prioritize sourcing from nodes with broad but short supply networks rather than long, narrow ones. Additionally, incorporating nodes with multiple parents significantly reduces $r^*$, as their production capacity must be distributed among multiple sources, decreasing overall efficiency.

Several promising extensions of this model merit future research. First, making a node’s maximum production capacity dependent on its number of parents could reveal additional topological effects. Second, replacing the fixed demand rate $r$ with a stochastic variable would allow for the study of demand fluctuations. Finally, introducing correlations between node production capacities—such as shared energy costs—could better reflect real-world dependencies. A control parameter could then adjust these correlations, enabling an analysis of their impact on system criticality.\\

\textit{Acknowledgements---}YF is funded by the Convergence Institute CLand
\url{https://cland.lsce.ipsl.fr}. YF also thanks Yvonne Feld for her
help with the schematic depiction of our model.

\bibliography{demand_paper.bib}

\begin{thebibliography}{20}%
\makeatletter
\providecommand \@ifxundefined [1]{%
 \@ifx{#1\undefined}
}%
\providecommand \@ifnum [1]{%
 \ifnum #1\expandafter \@firstoftwo
 \else \expandafter \@secondoftwo
 \fi
}%
\providecommand \@ifx [1]{%
 \ifx #1\expandafter \@firstoftwo
 \else \expandafter \@secondoftwo
 \fi
}%
\providecommand \natexlab [1]{#1}%
\providecommand \enquote  [1]{``#1''}%
\providecommand \bibnamefont  [1]{#1}%
\providecommand \bibfnamefont [1]{#1}%
\providecommand \citenamefont [1]{#1}%
\providecommand \href@noop [0]{\@secondoftwo}%
\providecommand \href [0]{\begingroup \@sanitize@url \@href}%
\providecommand \@href[1]{\@@startlink{#1}\@@href}%
\providecommand \@@href[1]{\endgroup#1\@@endlink}%
\providecommand \@sanitize@url [0]{\catcode `\\12\catcode `\$12\catcode
  `\&12\catcode `\#12\catcode `\^12\catcode `\_12\catcode `\%12\relax}%
\providecommand \@@startlink[1]{}%
\providecommand \@@endlink[0]{}%
\providecommand \url  [0]{\begingroup\@sanitize@url \@url }%
\providecommand \@url [1]{\endgroup\@href {#1}{\urlprefix }}%
\providecommand \urlprefix  [0]{URL }%
\providecommand \Eprint [0]{\href }%
\providecommand \doibase [0]{https://doi.org/}%
\providecommand \selectlanguage [0]{\@gobble}%
\providecommand \bibinfo  [0]{\@secondoftwo}%
\providecommand \bibfield  [0]{\@secondoftwo}%
\providecommand \translation [1]{[#1]}%
\providecommand \BibitemOpen [0]{}%
\providecommand \bibitemStop [0]{}%
\providecommand \bibitemNoStop [0]{.\EOS\space}%
\providecommand \EOS [0]{\spacefactor3000\relax}%
\providecommand \BibitemShut  [1]{\csname bibitem#1\endcsname}%
\let\auto@bib@innerbib\@empty
\bibitem [{\citenamefont {Nagurney}\ \emph {et~al.}(2013)\citenamefont
  {Nagurney}, \citenamefont {Yu}, \citenamefont {Masoumi},\ and\ \citenamefont
  {Nagurney}}]{nagurney2013}%
  \BibitemOpen
  \bibfield  {author} {\bibinfo {author} {\bibfnamefont {A.}~\bibnamefont
  {Nagurney}}, \bibinfo {author} {\bibfnamefont {M.}~\bibnamefont {Yu}},
  \bibinfo {author} {\bibfnamefont {A.~H.}\ \bibnamefont {Masoumi}},\ and\
  \bibinfo {author} {\bibfnamefont {L.~S.}\ \bibnamefont {Nagurney}},\ }\href
  {https://doi.org/10.1007/978-1-4614-6277-4} {\emph {\bibinfo {title}
  {Networks Against Time}}},\ \bibinfo {edition} {1st}\ ed.\ (\bibinfo
  {publisher} {Springer New York, NY},\ \bibinfo {year} {2013})\BibitemShut
  {NoStop}%
\bibitem [{\citenamefont {Nagurney}(2023)}]{nagurney2023}%
  \BibitemOpen
  \bibfield  {author} {\bibinfo {author} {\bibfnamefont {A.}~\bibnamefont
  {Nagurney}},\ }\href {https://doi.org/10.1007/978-3-031-20855-3} {\emph
  {\bibinfo {title} {Labor and Supply Chain Networks}}},\ \bibinfo {edition}
  {1st}\ ed.\ (\bibinfo  {publisher} {Springer Cham},\ \bibinfo {year}
  {2023})\BibitemShut {NoStop}%
\bibitem [{\citenamefont {Nagurney}\ \emph {et~al.}(2024)\citenamefont
  {Nagurney}, \citenamefont {Hassani}, \citenamefont {Nivievskyi},\ and\
  \citenamefont {Martyshev}}]{nagurney2024}%
  \BibitemOpen
  \bibfield  {author} {\bibinfo {author} {\bibfnamefont {A.}~\bibnamefont
  {Nagurney}}, \bibinfo {author} {\bibfnamefont {D.}~\bibnamefont {Hassani}},
  \bibinfo {author} {\bibfnamefont {O.}~\bibnamefont {Nivievskyi}},\ and\
  \bibinfo {author} {\bibfnamefont {P.}~\bibnamefont {Martyshev}},\ }\bibfield
  {title} {\bibinfo {title} {Multicommodity international agricultural trade
  network equilibrium: Competition for limited production and transportation
  capacity under disaster scenarios with implications for food security},\
  }\href {https://doi.org/10.1016/j.ejor.2023.11.010} {\bibfield  {journal}
  {\bibinfo  {journal} {European Journal of Operational Research}\ }\textbf
  {\bibinfo {volume} {314}},\ \bibinfo {pages} {1127} (\bibinfo {year}
  {2024})}\BibitemShut {NoStop}%
\bibitem [{\citenamefont {Bartezzaghi}\ \emph {et~al.}(2016)\citenamefont
  {Bartezzaghi}, \citenamefont {Cagliano}, \citenamefont {Caniato},\ and\
  \citenamefont {Ronchi}}]{Bartezzaghi2016}%
  \BibitemOpen
  \bibinfo {editor} {\bibfnamefont {E.}~\bibnamefont {Bartezzaghi}}, \bibinfo
  {editor} {\bibfnamefont {R.}~\bibnamefont {Cagliano}}, \bibinfo {editor}
  {\bibfnamefont {F.}~\bibnamefont {Caniato}},\ and\ \bibinfo {editor}
  {\bibfnamefont {S.}~\bibnamefont {Ronchi}},\ eds.,\ \href
  {https://doi.org/10.1007/978-3-319-31104-3} {\emph {\bibinfo {title} {A
  Journey through Manufacturing and Supply Chain Strategy Research}}}\
  (\bibinfo  {publisher} {Springer Cham},\ \bibinfo {year} {2016})\BibitemShut
  {NoStop}%
\bibitem [{\citenamefont {Katsaliaki}\ \emph {et~al.}(2022)\citenamefont
  {Katsaliaki}, \citenamefont {Galetsi},\ and\ \citenamefont
  {Kumar}}]{Katsaliaki2022}%
  \BibitemOpen
  \bibfield  {author} {\bibinfo {author} {\bibfnamefont {K.}~\bibnamefont
  {Katsaliaki}}, \bibinfo {author} {\bibfnamefont {P.}~\bibnamefont
  {Galetsi}},\ and\ \bibinfo {author} {\bibfnamefont {S.}~\bibnamefont
  {Kumar}},\ }\bibfield  {title} {\bibinfo {title} {Supply chain disruptions
  and resilience: a major review and future research agenda},\ }\href
  {https://doi.org/10.1007/s10479-020-03912-1} {\bibfield  {journal} {\bibinfo
  {journal} {Annals of Operations Research}\ }\textbf {\bibinfo {volume}
  {319}},\ \bibinfo {pages} {965} (\bibinfo {year} {2022})}\BibitemShut
  {NoStop}%
\bibitem [{\citenamefont {Dong}\ \emph {et~al.}(2004)\citenamefont {Dong},
  \citenamefont {Zhang},\ and\ \citenamefont {Nagurney}}]{dong2004}%
  \BibitemOpen
  \bibfield  {author} {\bibinfo {author} {\bibfnamefont {J.}~\bibnamefont
  {Dong}}, \bibinfo {author} {\bibfnamefont {D.}~\bibnamefont {Zhang}},\ and\
  \bibinfo {author} {\bibfnamefont {A.}~\bibnamefont {Nagurney}},\ }\bibfield
  {title} {\bibinfo {title} {A supply chain network equilibrium model with
  random demands},\ }\href {https://doi.org/10.1016/S0377-2217(03)00023-7}
  {\bibfield  {journal} {\bibinfo  {journal} {European Journal of Operational
  Research}\ }\textbf {\bibinfo {volume} {156}},\ \bibinfo {pages} {194}
  (\bibinfo {year} {2004})},\ \bibinfo {note} {eURO Excellence in Practice
  Award 2001}\BibitemShut {NoStop}%
\bibitem [{\citenamefont {Colon}\ and\ \citenamefont {Ghil}(2017)}]{Colon2017}%
  \BibitemOpen
  \bibfield  {author} {\bibinfo {author} {\bibfnamefont {C.}~\bibnamefont
  {Colon}}\ and\ \bibinfo {author} {\bibfnamefont {M.}~\bibnamefont {Ghil}},\
  }\bibfield  {title} {\bibinfo {title} {{Economic networks:
  Heterogeneity-induced vulnerability and loss of synchronization}},\ }\href
  {https://doi.org/10.1063/1.5017851} {\bibfield  {journal} {\bibinfo
  {journal} {Chaos: An Interdisciplinary Journal of Nonlinear Science}\
  }\textbf {\bibinfo {volume} {27}},\ \bibinfo {pages} {126703} (\bibinfo
  {year} {2017})}\BibitemShut {NoStop}%
\bibitem [{\citenamefont {Caraiani}\ \emph {et~al.}(2024)\citenamefont
  {Caraiani}, \citenamefont {Dima}, \citenamefont {Păun}, \citenamefont
  {Stamule},\ and\ \citenamefont {Vargas}}]{Caraiani2024}%
  \BibitemOpen
  \bibfield  {author} {\bibinfo {author} {\bibfnamefont {P.}~\bibnamefont
  {Caraiani}}, \bibinfo {author} {\bibfnamefont {A.~M.}\ \bibnamefont {Dima}},
  \bibinfo {author} {\bibfnamefont {C.}~\bibnamefont {Păun}}, \bibinfo
  {author} {\bibfnamefont {T.}~\bibnamefont {Stamule}},\ and\ \bibinfo {author}
  {\bibfnamefont {M.~V.}\ \bibnamefont {Vargas}},\ }\bibfield  {title}
  {\bibinfo {title} {Production networks and resilience: How dense production
  networks shield economies in financial crisis},\ }\href
  {https://doi.org/10.1371/journal.pone.0302012} {\bibfield  {journal}
  {\bibinfo  {journal} {PLOS ONE}\ }\textbf {\bibinfo {volume} {19}},\ \bibinfo
  {pages} {1} (\bibinfo {year} {2024})}\BibitemShut {NoStop}%
\bibitem [{\citenamefont {Battiston}\ \emph {et~al.}(2007)\citenamefont
  {Battiston}, \citenamefont {{Delli Gatti}}, \citenamefont {Gallegati},
  \citenamefont {Greenwald},\ and\ \citenamefont {Stiglitz}}]{battiston2007}%
  \BibitemOpen
  \bibfield  {author} {\bibinfo {author} {\bibfnamefont {S.}~\bibnamefont
  {Battiston}}, \bibinfo {author} {\bibfnamefont {D.}~\bibnamefont {{Delli
  Gatti}}}, \bibinfo {author} {\bibfnamefont {M.}~\bibnamefont {Gallegati}},
  \bibinfo {author} {\bibfnamefont {B.}~\bibnamefont {Greenwald}},\ and\
  \bibinfo {author} {\bibfnamefont {J.~E.}\ \bibnamefont {Stiglitz}},\
  }\bibfield  {title} {\bibinfo {title} {Credit chains and bankruptcy
  propagation in production networks},\ }\href
  {https://doi.org/10.1016/j.jedc.2007.01.004} {\bibfield  {journal} {\bibinfo
  {journal} {Journal of Economic Dynamics and Control}\ }\textbf {\bibinfo
  {volume} {31}},\ \bibinfo {pages} {2061} (\bibinfo {year} {2007})},\ \bibinfo
  {note} {tenth Workshop on Economic Heterogeneous Interacting
  Agents}\BibitemShut {NoStop}%
\bibitem [{\citenamefont {Hearnshaw}\ and\ \citenamefont
  {Wilson}(2013)}]{Hearnshaw2013}%
  \BibitemOpen
  \bibfield  {author} {\bibinfo {author} {\bibfnamefont {E.~J.}\ \bibnamefont
  {Hearnshaw}}\ and\ \bibinfo {author} {\bibfnamefont {M.~M.}\ \bibnamefont
  {Wilson}},\ }\bibfield  {title} {\bibinfo {title} {A complex network approach
  to supply chain network theory},\ }\href
  {https://doi.org/10.1108/01443571311307343} {\bibfield  {journal} {\bibinfo
  {journal} {International Journal of Operations {\&} Production Management}\
  }\textbf {\bibinfo {volume} {33}},\ \bibinfo {pages} {442} (\bibinfo {year}
  {2013})}\BibitemShut {NoStop}%
\bibitem [{\citenamefont {Ramanathan}\ and\ \citenamefont
  {Ramanathan}(2020)}]{Ramanathan2020}%
  \BibitemOpen
  \bibinfo {editor} {\bibfnamefont {U.}~\bibnamefont {Ramanathan}}\ and\
  \bibinfo {editor} {\bibfnamefont {R.}~\bibnamefont {Ramanathan}},\ eds.,\
  \href {https://doi.org/10.1007/978-3-030-48876-5} {\emph {\bibinfo {title}
  {Sustainable Supply Chains: Strategies, Issues, and Models}}},\ \bibinfo
  {edition} {1st}\ ed.\ (\bibinfo  {publisher} {Springer Cham},\ \bibinfo
  {year} {2020})\BibitemShut {NoStop}%
\bibitem [{\citenamefont {Sawik}(2024)}]{sawik2024}%
  \BibitemOpen
  \bibfield  {author} {\bibinfo {author} {\bibfnamefont {T.}~\bibnamefont
  {Sawik}},\ }\href {https://doi.org/10.1007/978-3-031-57927-1} {\emph
  {\bibinfo {title} {Stochastic Programming in Supply Chain Risk Management}}}\
  (\bibinfo  {publisher} {Springer Cham},\ \bibinfo {year} {2024})\BibitemShut
  {NoStop}%
\bibitem [{\citenamefont {Moran}\ \emph {et~al.}(2024)\citenamefont {Moran},
  \citenamefont {Romeijnders}, \citenamefont {Doussal}, \citenamefont
  {Pijpers}, \citenamefont {Weitzel}, \citenamefont {Panja},\ and\
  \citenamefont {Bouchaud}}]{moran2024}%
  \BibitemOpen
  \bibfield  {author} {\bibinfo {author} {\bibfnamefont {J.}~\bibnamefont
  {Moran}}, \bibinfo {author} {\bibfnamefont {M.}~\bibnamefont {Romeijnders}},
  \bibinfo {author} {\bibfnamefont {P.~L.}\ \bibnamefont {Doussal}}, \bibinfo
  {author} {\bibfnamefont {F.~P.}\ \bibnamefont {Pijpers}}, \bibinfo {author}
  {\bibfnamefont {U.}~\bibnamefont {Weitzel}}, \bibinfo {author} {\bibfnamefont
  {D.}~\bibnamefont {Panja}},\ and\ \bibinfo {author} {\bibfnamefont {J.-P.}\
  \bibnamefont {Bouchaud}},\ }\bibfield  {title} {\bibinfo {title} {Timeliness
  criticality in complex systems},\ }\bibfield  {journal} {\bibinfo  {journal}
  {Nature Physics}\ }\href {https://doi.org/10.1038/s41567-024-02525-w}
  {10.1038/s41567-024-02525-w} (\bibinfo {year} {2024})\BibitemShut {NoStop}%
\bibitem [{\citenamefont {Meixell}\ and\ \citenamefont
  {Gargeya}(2005)}]{Meixell2005}%
  \BibitemOpen
  \bibfield  {author} {\bibinfo {author} {\bibfnamefont {M.~J.}\ \bibnamefont
  {Meixell}}\ and\ \bibinfo {author} {\bibfnamefont {V.~B.}\ \bibnamefont
  {Gargeya}},\ }\bibfield  {title} {\bibinfo {title} {Global supply chain
  design: A literature review and critique},\ }\href
  {https://doi.org/https://doi.org/10.1016/j.tre.2005.06.003} {\bibfield
  {journal} {\bibinfo  {journal} {Transportation Research Part E: Logistics and
  Transportation Review}\ }\textbf {\bibinfo {volume} {41}},\ \bibinfo {pages}
  {531} (\bibinfo {year} {2005})}\BibitemShut {NoStop}%
\bibitem [{\citenamefont {Wiedmer}\ and\ \citenamefont
  {Griffis}(2021)}]{Wiedmer2021}%
  \BibitemOpen
  \bibfield  {author} {\bibinfo {author} {\bibfnamefont {R.}~\bibnamefont
  {Wiedmer}}\ and\ \bibinfo {author} {\bibfnamefont {S.~E.}\ \bibnamefont
  {Griffis}},\ }\bibfield  {title} {\bibinfo {title} {Structural
  characteristics of complex supply chain networks},\ }\href
  {https://doi.org/10.1111/jbl.12283} {\bibfield  {journal} {\bibinfo
  {journal} {Journal of Business Logistics}\ }\textbf {\bibinfo {volume}
  {42}},\ \bibinfo {pages} {264} (\bibinfo {year} {2021})}\BibitemShut
  {NoStop}%
\bibitem [{\citenamefont {Kito}\ \emph {et~al.}(2014)\citenamefont {Kito},
  \citenamefont {Brintrup}, \citenamefont {New},\ and\ \citenamefont
  {Reed-Tsochas}}]{Kito2014}%
  \BibitemOpen
  \bibfield  {author} {\bibinfo {author} {\bibfnamefont {T.}~\bibnamefont
  {Kito}}, \bibinfo {author} {\bibfnamefont {A.}~\bibnamefont {Brintrup}},
  \bibinfo {author} {\bibfnamefont {S.}~\bibnamefont {New}},\ and\ \bibinfo
  {author} {\bibfnamefont {F.}~\bibnamefont {Reed-Tsochas}},\ }\bibfield
  {title} {\bibinfo {title} {The structure of the toyota supply network: An
  empirical analysis},\ }\bibfield  {journal} {\bibinfo  {journal} {Saïd
  Business School WP 2014-3}\ }\href {https://doi.org/10.2139/ssrn.2412512}
  {10.2139/ssrn.2412512} (\bibinfo {year} {2014})\BibitemShut {NoStop}%
\bibitem [{\citenamefont {Perera}\ \emph {et~al.}(2018)\citenamefont {Perera},
  \citenamefont {Bell}, \citenamefont {Piraveenan}, \citenamefont
  {Kasthurirathna},\ and\ \citenamefont {Parhi}}]{Perera2018}%
  \BibitemOpen
  \bibfield  {author} {\bibinfo {author} {\bibfnamefont {S.~S.}\ \bibnamefont
  {Perera}}, \bibinfo {author} {\bibfnamefont {M.~G.~H.}\ \bibnamefont {Bell}},
  \bibinfo {author} {\bibfnamefont {M.}~\bibnamefont {Piraveenan}}, \bibinfo
  {author} {\bibfnamefont {D.}~\bibnamefont {Kasthurirathna}},\ and\ \bibinfo
  {author} {\bibfnamefont {M.}~\bibnamefont {Parhi}},\ }\bibfield  {title}
  {\bibinfo {title} {Topological structure of manufacturing industry supply
  chain networks},\ }\href {https://doi.org/10.1155/2018/3924361} {\bibfield
  {journal} {\bibinfo  {journal} {Complexity}\ }\textbf {\bibinfo {volume}
  {2018}},\ \bibinfo {pages} {3924361} (\bibinfo {year} {2018})}\BibitemShut
  {NoStop}%
\bibitem [{\citenamefont {Fleischmann}\ \emph {et~al.}(2000)\citenamefont
  {Fleischmann}, \citenamefont {Krikke}, \citenamefont {Dekker},\ and\
  \citenamefont {Flapper}}]{Fleischmann2000}%
  \BibitemOpen
  \bibfield  {author} {\bibinfo {author} {\bibfnamefont {M.}~\bibnamefont
  {Fleischmann}}, \bibinfo {author} {\bibfnamefont {H.~R.}\ \bibnamefont
  {Krikke}}, \bibinfo {author} {\bibfnamefont {R.}~\bibnamefont {Dekker}},\
  and\ \bibinfo {author} {\bibfnamefont {S.~D.~P.}\ \bibnamefont {Flapper}},\
  }\bibfield  {title} {\bibinfo {title} {A characterisation of logistics
  networks for product recovery},\ }\href
  {https://doi.org/https://doi.org/10.1016/S0305-0483(00)00022-0} {\bibfield
  {journal} {\bibinfo  {journal} {Omega}\ }\textbf {\bibinfo {volume} {28}},\
  \bibinfo {pages} {653} (\bibinfo {year} {2000})}\BibitemShut {NoStop}%
\bibitem [{\citenamefont {Francas}\ and\ \citenamefont
  {Minner}(2009)}]{Francas2008}%
  \BibitemOpen
  \bibfield  {author} {\bibinfo {author} {\bibfnamefont {D.}~\bibnamefont
  {Francas}}\ and\ \bibinfo {author} {\bibfnamefont {S.}~\bibnamefont
  {Minner}},\ }\bibfield  {title} {\bibinfo {title} {Manufacturing network
  configuration in supply chains with product recovery},\ }\href
  {https://doi.org/https://doi.org/10.1016/j.omega.2008.07.007} {\bibfield
  {journal} {\bibinfo  {journal} {Omega}\ }\textbf {\bibinfo {volume} {37}},\
  \bibinfo {pages} {757} (\bibinfo {year} {2009})},\ \bibinfo {note} {role of
  Flexibility in Supply Chain Design and Modeling}\BibitemShut {NoStop}%
\bibitem [{\citenamefont {Hartmann}(2015)}]{hartmann2015}%
  \BibitemOpen
  \bibfield  {author} {\bibinfo {author} {\bibfnamefont {A.~K.}\ \bibnamefont
  {Hartmann}},\ }\href {https://doi.org/10.1142/9019} {\emph {\bibinfo {title}
  {{Big Practical Guide to Computer Simulations}}}},\ \bibinfo {edition} {2nd}\
  ed.\ (\bibinfo  {publisher} {World Scientific},\ \bibinfo {year}
  {2015})\BibitemShut {NoStop}%
\end{thebibliography}%


\begin{thebibliography}{4}%
\makeatletter
\providecommand \@ifxundefined [1]{%
 \@ifx{#1\undefined}
}%
\providecommand \@ifnum [1]{%
 \ifnum #1\expandafter \@firstoftwo
 \else \expandafter \@secondoftwo
 \fi
}%
\providecommand \@ifx [1]{%
 \ifx #1\expandafter \@firstoftwo
 \else \expandafter \@secondoftwo
 \fi
}%
\providecommand \natexlab [1]{#1}%
\providecommand \enquote  [1]{``#1''}%
\providecommand \bibnamefont  [1]{#1}%
\providecommand \bibfnamefont [1]{#1}%
\providecommand \citenamefont [1]{#1}%
\providecommand \href@noop [0]{\@secondoftwo}%
\providecommand \href [0]{\begingroup \@sanitize@url \@href}%
\providecommand \@href[1]{\@@startlink{#1}\@@href}%
\providecommand \@@href[1]{\endgroup#1\@@endlink}%
\providecommand \@sanitize@url [0]{\catcode `\\12\catcode `\$12\catcode
  `\&12\catcode `\#12\catcode `\^12\catcode `\_12\catcode `\%12\relax}%
\providecommand \@@startlink[1]{}%
\providecommand \@@endlink[0]{}%
\providecommand \url  [0]{\begingroup\@sanitize@url \@url }%
\providecommand \@url [1]{\endgroup\@href {#1}{\urlprefix }}%
\providecommand \urlprefix  [0]{URL }%
\providecommand \Eprint [0]{\href }%
\providecommand \doibase [0]{https://doi.org/}%
\providecommand \selectlanguage [0]{\@gobble}%
\providecommand \bibinfo  [0]{\@secondoftwo}%
\providecommand \bibfield  [0]{\@secondoftwo}%
\providecommand \translation [1]{[#1]}%
\providecommand \BibitemOpen [0]{}%
\providecommand \bibitemStop [0]{}%
\providecommand \bibitemNoStop [0]{.\EOS\space}%
\providecommand \EOS [0]{\spacefactor3000\relax}%
\providecommand \BibitemShut  [1]{\csname bibitem#1\endcsname}%
\let\auto@bib@innerbib\@empty
\bibitem [{\citenamefont {Meixell}\ and\ \citenamefont
  {Gargeya}(2005)}]{Meixell2005}%
  \BibitemOpen
  \bibfield  {author} {\bibinfo {author} {\bibfnamefont {M.~J.}\ \bibnamefont
  {Meixell}}\ and\ \bibinfo {author} {\bibfnamefont {V.~B.}\ \bibnamefont
  {Gargeya}},\ }\bibfield  {title} {\bibinfo {title} {Global supply chain
  design: A literature review and critique},\ }\href
  {https://doi.org/https://doi.org/10.1016/j.tre.2005.06.003} {\bibfield
  {journal} {\bibinfo  {journal} {Transportation Research Part E: Logistics and
  Transportation Review}\ }\textbf {\bibinfo {volume} {41}},\ \bibinfo {pages}
  {531} (\bibinfo {year} {2005})}\BibitemShut {NoStop}%
\bibitem [{\citenamefont {Wiedmer}\ and\ \citenamefont
  {Griffis}(2021)}]{Wiedmer2021}%
  \BibitemOpen
  \bibfield  {author} {\bibinfo {author} {\bibfnamefont {R.}~\bibnamefont
  {Wiedmer}}\ and\ \bibinfo {author} {\bibfnamefont {S.~E.}\ \bibnamefont
  {Griffis}},\ }\bibfield  {title} {\bibinfo {title} {Structural
  characteristics of complex supply chain networks},\ }\href
  {https://doi.org/10.1111/jbl.12283} {\bibfield  {journal} {\bibinfo
  {journal} {Journal of Business Logistics}\ }\textbf {\bibinfo {volume}
  {42}},\ \bibinfo {pages} {264} (\bibinfo {year} {2021})}\BibitemShut
  {NoStop}%
\bibitem [{\citenamefont {Kito}\ \emph {et~al.}(2014)\citenamefont {Kito},
  \citenamefont {Brintrup}, \citenamefont {New},\ and\ \citenamefont
  {Reed-Tsochas}}]{kito2014}%
  \BibitemOpen
  \bibfield  {author} {\bibinfo {author} {\bibfnamefont {T.}~\bibnamefont
  {Kito}}, \bibinfo {author} {\bibfnamefont {A.}~\bibnamefont {Brintrup}},
  \bibinfo {author} {\bibfnamefont {S.}~\bibnamefont {New}},\ and\ \bibinfo
  {author} {\bibfnamefont {F.}~\bibnamefont {Reed-Tsochas}},\ }\bibfield
  {title} {\bibinfo {title} {The structure of the toyota supply network: An
  empirical analysis},\ }\bibfield  {journal} {\bibinfo  {journal} {Saïd
  Business School WP 2014-3}\ }\href {https://doi.org/10.2139/ssrn.2412512}
  {10.2139/ssrn.2412512} (\bibinfo {year} {2014})\BibitemShut {NoStop}%
\bibitem [{\citenamefont {Perera}\ \emph {et~al.}(2018)\citenamefont {Perera},
  \citenamefont {Bell}, \citenamefont {Piraveenan}, \citenamefont
  {Kasthurirathna},\ and\ \citenamefont {Parhi}}]{Perera2018}%
  \BibitemOpen
  \bibfield  {author} {\bibinfo {author} {\bibfnamefont {S.~S.}\ \bibnamefont
  {Perera}}, \bibinfo {author} {\bibfnamefont {M.~G.~H.}\ \bibnamefont {Bell}},
  \bibinfo {author} {\bibfnamefont {M.}~\bibnamefont {Piraveenan}}, \bibinfo
  {author} {\bibfnamefont {D.}~\bibnamefont {Kasthurirathna}},\ and\ \bibinfo
  {author} {\bibfnamefont {M.}~\bibnamefont {Parhi}},\ }\bibfield  {title}
  {\bibinfo {title} {Topological structure of manufacturing industry supply
  chain networks},\ }\href {https://doi.org/10.1155/2018/3924361} {\bibfield
  {journal} {\bibinfo  {journal} {Complexity}\ }\textbf {\bibinfo {volume}
  {2018}},\ \bibinfo {pages} {3924361} (\bibinfo {year} {2018})}\BibitemShut
  {NoStop}%
\end{thebibliography}%

\end{document}


\title{Supplementary Material for: \\ Critical demand in a stochastic model of flows in supply networks}

\author{Yannick Feld}

\affiliation{Université Paris-Saclay, CNRS, CEA, Institut de
  Physique Théorique, 91191 Gif-sur-Yvette, France}
\author{Marc Barthelemy$^{1,}$}
\affiliation{Centre d'Analyse et de Math\'ematique Sociales (CNRS/EHESS) 54 Avenue de Raspail, 75006 Paris, France}

\date{\today}

\maketitle

\section{Assumption of acyclic networks}

Nodes of supply chain networks 
are often described by categorizing them into different tiers \cite{Meixell2005},
for example manufacturer (tier 1), the suppliers of the manufacturer (tier 2), the suppliers of the tier 2 suppliers (tier 3)
and so on.
The overall assumption is that products or services flow from the higher tiers to the lower tiers,
while each layer adds some value by processing or assembling the inputs into new products.

According to Wiedmer et al.~\cite{Wiedmer2021}, who in 2021 investigated a dataset of 21 focus firms and their corresponding networks, 
with a total of about 185000 nodes, horizontal relationships between firms of the same tier are rare, which makes sense since 
firms of the same tier fulfill roughly the same role in the network and thus compete for resources.

Contrastingly a prior study by Kito et al.~\cite{kito2014} in 2014, who investigated a much smaller dataset,
namely the supply network of Toyota, 
found a significant number of ties between firms 
of the same tier. They noted that this demonstrates the fuzziness of the tier boundaries.

On the other hand, in 2018 Perera et al.~\cite{Perera2018} investigated multiple directed material flow networks,
i.e., networks that, as the name implies, describe only the flow of materials as opposed to 
also tracking other things like services or other cooperations. Within their dataset they found 
\emph{no} links between nodes of the 
same functional tier.

This leads us to a fundamental assumption. While some of the items required to produce the root item or 
an intermediate items may depend on common items (e.g. multiple products use identical screws),
 i.e., nodes may have common descendants or ancestors, 
we assume no node is its \emph{own} ancestor.
Thus, based on the aforementioned studies, we can approximate the production network 
well by describing it with directed acyclic graph,
i.e., a network where directed loops are forbidden, 
without the need to assign tiers to any of the nodes.

\section{Pseudo-code for solving the model}

Here we describe the algorithms used for solving the model.
First we describe an algorithm that determines the order in which we need to 
calculate the demand of the nodes in the supply network (see \autoref{RootOrder}).
The order is important, because, to calculate the demand of any node,
the demand of its parents needs to be known already. 

The algorithm begins by initializing a stack and an ordered list.
The ordered list will store the final sequence in which the demands are calculated,
while the stack is used to track the nodes that are ready to be processed,
i.e., for which all parents are already in the ordered list.
Additionally we maintain a vector of counters which is used to keep track of how many 
of a node's parents have already been processed.

Initially we push the root node onto the stack, as it has no parents.
Next we do the following until the stack is empty.
We pop a node from the stack and add it to the ordered list. 
Next we increment the parent counters of all of its children.
If this results in the parent counter of a child becoming equal to the 
actual number of parents that said child has, we push the child on the stack,
indicating that it is ready to be processed.

Once this algorithm completes, the ordered list will contain a sequence where, 
for each node, its parents appear earlier than itself in the list. 
This ensures that the following demand calculations respect the dependencies of the network.
Since the networks we look at are static, 
this order only needs to be calculated once for each network.

\begin{algorithm}[h]
    \setstretch{1}
    \CommentSty{\color{crimson}}
    \caption{Calculate order for demands}\label{RootOrder}
        \SetKwInOut{Input}{input}
        \SetKwInOut{Output}{output}
        \Input{Nodes\tcp*{Vector of nodes. This is the network. Nodes contain indices corresponding to their parents and children. Node 0 is the root node}}
        \Output{Ordered List of Nodes for calculating demand}
        \SetKwBlock{Beginn}{beginn}{ende}
        \Begin{
            \textbf{Initialize} stack\;
            \textbf{push} 0 on stack\tcp*{Push the root node on the stack}
            \textbf{Initialize} ordered\_list\;
            \textbf{Initialize} parent\_count\_vec \tcp*{vector of same length as Nodes, initially contains only zeros}
            \While{\emph{stack not empty}}{
                current\_node\_idx = stack.pop()\tcp*{Take last element from stack}
                \textbf{push} current\_node\_idx on ordered\_list\;
                \For{\emph{child\_idx in Nodes[current\_node\_idx].children}}{
                    parent\_count\_vec [child\_idx] += 1\;
                    \tcp{Check if all parents of the child are already in the list}
                    \If{\emph{parent\_count\_vec[child\_idx] == Nodes[child\_idx].parent\_count}}{
                        \textbf{push} child\_idx on stack
                    }
                }
            }
            \Return ordered\_list
        }
\end{algorithm}

The next algorithm is used to calculate the actual demands (see \autoref{DemandCalc}).
At first we initialize a demand vector, where each entry represents 
the demand of a node. All demands are initialized with zero.
Next, we calculate the new root demand by using the unsatisfied demand 
from the last time step $u(t-1)$ and the demand rate $r$.
Then we iterate through the ordered list created with the previous algorithm.
For each node $i$ we iterate over its children $l$ 
and calculate how much needs to be ordered from them, i.e.,
we calculate $\Omega_{il}$, and add that to the demand value of the respective child.

\begin{algorithm}[h]
    \setstretch{1}
    \CommentSty{\color{blue}}
    \caption{Calculate demands}\label{DemandCalc}
        \SetKwInOut{Input}{input}
        \SetKwInOut{Output}{output}
        \Input{$u(t-1)$, ordered\_list, $r$, Nodes, k(t)}
        \Output{D, $\Omega$}
        \SetKwBlock{Beginn}{beginn}{ende}
        \Begin{
            \textbf{Initialize} D\tcp*{vector that initially contains a 0 for each node. This is the demand, see Eq.~3 (main text)}
            \textbf{Initialize} $\Omega$\tcp*{$N \times N$ matrix that initially contains only zeros. See Eq.~2 (main text)}
            D[0] = $u(t-1)$ + $r$ \tcp*{Update demand of root node}
            \For{\emph{current\_idx in ordered\_list}}{
                \For{\emph{child\_idx in Nodes[current\_idx].children}}{
                    $\Omega$[current\_idx][child\_idx] = max(0, D[current\_idx] - k(t)[child\_idx][current\_idx])\;
                    D[child\_idx] += $\Omega$[current\_idx][child\_idx]\;
                }
            }
            \Return D, $\Omega$\;
        }
\end{algorithm}

The next algorithm is used to calculate the production $I$ for each node and calculate the stocks $k(t+1)$ for the next time step (see \autoref{ProductionCalc}).
To ensure that a node's children are processed before it, we reverse the ordered list from the previous 
algorithm, because it already provides the correct order.

We begin by initializing a zeroed vector $I$ for the production of each node and a zeroed matrix $a$ 
for the amount of product delivered during the current time step.
Then we iterate over all nodes in the reversed list, where the first node is guaranteed to be a leaf, i.e.,
a node without children.
For each node $i$ we draw a random value $m_i$ uniformly from $[0,1]$ to set its maximal production capacity.
The value $m_i$ is also the initial value for $I_i$ and if $i$ is a leaf, this is also the final value for the 
production.
Otherwise we iterate over all of $i$'s children $l$ to calculate the amount of product available from each child,
which is the sum of its current stock $k_{il}(t)$ and the amount delivered during the current step $a_{il}$.
This may further limit the production $I_i$.

Once the final value of $I_i$ is obtained, we iterate over all children again to calculate the stocks 
for the next time step $k(t+1)$.
Finally we distribute the production of node $i$ by  
updating the matrix $a$, dividing the output among its parents.

At the end of the algorithm we have calculated the production $I$ for all nodes and acquired the 
stock values for the next time step $k(t+1)$.

\begin{algorithm}[h]
    \setstretch{1}
    \CommentSty{\color{blue}}
    \caption{Calculate Production and stocks}\label{ProductionCalc}
        \SetKwInOut{Input}{input}
        \SetKwInOut{Output}{output}
        \Input{k(t), D, Nodes, reverse\_ordered\_list, $s$, $\Omega$}
        \Output{k(t+1), I}
        \SetKwBlock{Beginn}{beginn}{ende}
        \Begin{
            \textbf{Initialize} I\tcp*{vector that contains a 0 for each node. This is the production}
            \textbf{Initialize} a\tcp*{$N \times N$ matrix, initialized with zeros, see Eq.~5 (main text)}
            \For{\emph{current\_idx in reverse\_ordered\_list}}{
                $m$ = random uniform number in [0,1]\;
                I[current\_idx] = min(D[current\_idx], $m$)\;
                \tcp{Calculating production}
                \For{\emph{child\_idx in Nodes[current\_idx].children}}{
                    available\_child\_product = a[current\_idx][child\_idx] + k(t)[current\_idx][child\_idx]\;
                    I[current\_idx] = min(available\_child\_product, I[current\_idx])\;
                }
                \tcp{Calculating Stocks}
                \For{\emph{child\_idx in Nodes[current\_idx].children}}{
                    available\_child\_product = a[current\_idx][child\_idx] + k(t)[current\_idx][child\_idx]\;
                    k(t+1)[current\_idx][child\_idx] = min($s$, available\_child\_product - I[current\_idx])\;
                }
                \If{\emph{D[current\_idx] != 0}}{
                    \For{\emph{parent\_idx in Nodes[current\_idx].parents}}{
                        a[parent\_idx][current\_idx] = I[current\_idx] * $\Omega$[parent\_idx][current\_idx] / D[current\_idx]\;
                    }
                }
            }
            \Return k(t+1), I\;
        }
\end{algorithm}

\section{Measuring $r^*$}

We are interested in the critical root demand rate $r^*$ of our system, which is equal to the average amount 
of product that our system can produce when it is not limited by low demands.
To  measure $r^*$ we can measure $u(t)$ for some large $t$.
We have, in the limit of large $t$,
\begin{equation}
    \frac{u(t)}{t} =\max \left(0, r-r^* \right)~,
\end{equation}
which can be used to measure $r^*$ by fitting a line to $\frac{u(t)}{t}$ for large enough $t$.

\begin{figure}[htb]
    \centering
    \includegraphics[width=0.5\linewidth]{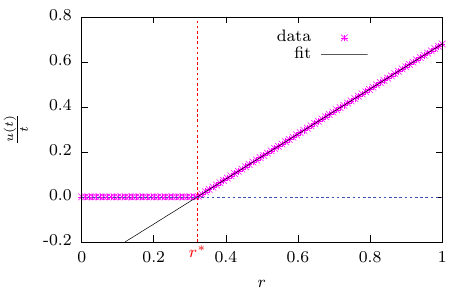} 
    \caption{\label{m_crit_example_fig} Example for measuring the critical root demand rate $r^*$ for $s=1$ and $N=21$.
    The nodes were configured into two chains of 10 nodes each, which then lead into the root node.}
\end{figure}

An example of this is shown in \autoref{m_crit_example_fig}.

\subsection{No stock}

\begin{figure}[htb]
    \includegraphics[width=0.5\linewidth]{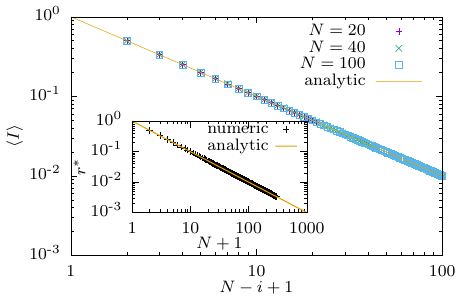} 
    \caption{\label{no_stock_fig} The outer plot shows the average production $\left<I\right>$ over $N+1-i$, while the inset 
    shows the critical root demand $r^*$ over $N+1$, both for $s=0$.}
\end{figure}

As mentioned in the main text, for $s=0$ the problem simplifies and we are able to give 
analytic solutions for the case of a simple chain. Here we compare those analytic results with numeric measurements. 
This is shown in \autoref{no_stock_fig} and we see that the numerics match the analytic prediction perfectly,
as expected.

When there are no stocks, the critical root demand $r^*$ is given by the average
production of the root node $i=0$, which implies that
\begin{equation}
    \label{No_stock_crit_eq}
    r^*=\frac{1}{N+1}~.
  \end{equation}
 This result is valid for any network where each node has at most one parent, i.e., 
for all tree-like networks. Additionally, for any value of $s$, \autoref{No_stock_crit_eq} also describes the average critical root demand 
for the quenched case where the $m_i$ are drawn in the beginning
and not changed afterwards. 
In this case the stocks $k_{ij}$ become irrelevant because there are no fluctuations in the supply of each node,
i.e., for large demands we have $a_{ij} =\text{const}$ and $I_i = \text{const}$. 
So there is either \emph{always} a surplus or  \emph{always} a shortage of a specific product and in neither of those cases does the stock 
have any effect, rendering $s$ irrelevant. Also, the only topological structure that is relevant in the quenched case 
or for $s=0$ is when any node $j$ has more than one 
parent. 
The demand of those nodes $j$ will be the sum of the demand of it's parents, i.e., larger than the demand of each individual parent
and the production has to be split 
between its parents. In this case one can still easily give an analytic solution for the 
critical root demand rate for any given network.

\section{Analytics for the 1d chain}

In the case of the simple chain each node has at most one child.
Since we only have a chain, we will, in the following, call $k_{ij} \to k_{i}$ and $a_{ij} \to a_i$,
as the second index would be redundant.
Node $i=N-1$ does not have stock, since it is the leaf node. Thus the first node for which the stock is of interest
is $i=N-2$.

We will now analytically determine the probability density $\rho (k_{N-2})$.
We assume the system to be above criticality, such that for long times we can assume the demand 
of each node to be arbitrarily large, i.e., the system is not limited by the demand in Eq.~(4).

Note that for the chain we can simply follow from the dynamics:
\begin{equation}
D_0\geq D_1 \geq … \geq D_N \geq D_0 - s(N-1)
\end{equation}

Node $N-1$ is our only leaf node. Since we assume our demand to be large, for the leaf Eq.~4 (main text) becomes:
\begin{equation}
    \label{eq_leaf_chain_m}
    I_{N-1}(t)=m_{N-1}(t)
\end{equation}

Since we only have a chain, Eq.~3 and Eq.~5 from the main text become:

\begin{align}
    \label{ai_eq_helper}
    a_i(t) &= I_{i+1}(t)\\
    \Rightarrow a_{N-2}(t) &= I_{N-1}(t) = m_{N-1}(t)
\end{align}

For node $N-2$ we have:
\begin{align}
    I_{N-2}(t) &= \min \left(m_{N-2}(t), a_{N-2}(t) + k_{N-2}(t), d_{N-2}(t)\right)\\
    &=\min \left(m_{N-2}(t), I_{N-1}(t) + k_{N-2}(t)\right)\\
    \label{I_n_2_needed_later_eq}
    &=\min \left(m_{N-2}(t), m_{N-1}(t) + k_{N-2}(t)\right)
\end{align}

Now lets look at the quantity that we are currently interested in by using Eq.~6 from the main text as well as 
\autoref{eq_leaf_chain_m} and \autoref{ai_eq_helper}:
\begin{align}
    k_{N-2}(t+1)&=\min[s,m_{N-2}(t)+k_{N-2}(t)-\min \{m_{N-1}(t),m_{N-1}(t+1) + k_{N-2}(t)\}]\\
    \label{jump_helper}
        &= \min \left[s, \max\left\{0,m_{N-2}(t+1)-m_{N-1}(t+1)+k_{N-2}(t)\right\}\right]
\end{align}

We now define
\begin{equation}
    v(t) := m_{N-2}(t)-m_{N-1}(t)~.
\end{equation}
And since all $m$ are uniformly distributed in $[0,1]$, this means the new variable $v$ is 
distributed according to
\begin{align}
    \label{veq}
    \rho_\nu(\nu) = \begin{cases} 1+\nu & \text{for}~-1 \leq \nu < 0\\
        1-\nu & \text{for}~0 \leq \nu \leq 1\\
        0 & \text{otherwise} 
    \end{cases}~.
\end{align}

Now let us look at an ensemble of our model, so that we can go to the master equation.
If we look at large times and assume to be at equilibrium, i.e., our distribution does not change anymore,
then we have 
\begin{equation}
    \label{master_eq}
    \rho^k_{N-2}(k) = \int_0^s \rho^k_{N-2}(\tilde{k}) J(\tilde{k} \to k) d\tilde{k}
\end{equation}
where $J(\tilde{k} \to k)$ is the jump-probability, i.e., the probability that at time $t$ the stock 
is at $k$ given 
that it was at  $\tilde{k}$ at time $t-1$. We later get this jump probability 
with the help of \autoref{jump_helper}.
Note that, since \autoref{veq} is symmetric around $0$, this means that the probability distribution that 
we search for has to be symmetric around $\frac{s}{2}$.
Thus it makes sense to look at 
\begin{equation}
    \hat{k}_{N-2}(t) ={k}_{N-2}(t) - \frac{s}{2} 
\end{equation}
since we would rather have a symmetry at 0, to make the following a bit easier.

We know that the distribution we search for needs to exhibit two $\delta$-peaks 
at $\hat{k}= \pm \frac{s}{2}$, which is a result of the minimum and maximum in \autoref{jump_helper}.
And for symmetry reasons those delta peaks also need to be of the same height.
Therefore our first approach for the distribution is
\begin{align}
    \rho^{\hat{k}}_{N-2}(k) = \left(\delta\left(k-\frac{s}{2}\right) + \delta \left(k+\frac{s}{2}\right)\right) \zeta  + \hat{f}(k)
\end{align}
where $\zeta$ is some constant and $\hat{f}(k)$ is a yet to be determined function.
For $-\frac{s}{2}<\hat{k}<\frac{s}{2}$ we know 
\begin{equation}
    J(\tilde{k} \to k) = \rho_\nu(k-\tilde{k})
\end{equation}
and can thus use \autoref{master_eq}:
\begin{align}
    \hat{f}(k) &= \int_{-s/2}^{s/2} \rho^{\hat{k}}_{N-2}(\tilde{k})   \rho_{\nu}(k-\tilde{k}) d\tilde{k}
\end{align}
from now on we assume $s=1$
\begin{align}
    \hat{f}(k)&=\int_{-1/2}^{k} \rho^{\hat{k}}_{N-2}(\tilde{k}) (1-k+\tilde{k}) d\tilde{k} +\int_{k}^{1/2} \rho^{\hat{k}}_{N-2}(\tilde{k}) (1+k-\tilde{k}) d\tilde{k}\\
    &=… + \zeta \int_{-1/2}^{k} \delta \left(\tilde{k} + \frac{1}{2}\right) (1-k+\tilde{k}) d\tilde{k} 
    + \zeta \int_{k}^{1/2} \delta \left(\tilde{k} - \frac{1}{2}\right) (1+k-\tilde{k}) d\tilde{k}\\
    &=… + \zeta\\
    & = \int_{-1/2}^{k} \hat{f}(k) (1-k+\tilde{k}) d\tilde{k} +\int_{k}^{1/2} \hat{f}(k) (1+k-\tilde{k}) d\tilde{k} + \zeta
    \label{integral_to_solve}
\end{align}

So now we need a function $\hat{f}(k)$ for which the integrals result in $\hat{f}(k)-\zeta$.
Let's continue with an educated guess and assume
\begin{equation}
    \hat{f}(k)= \lambda \cos\left(\sqrt{2} k\right)
\end{equation}
where $\lambda$ is some constant.
Plugging this into \autoref{integral_to_solve} gives:
\begin{equation}
    \label{fk_equation}
    \hat{f}(k)=\zeta + \lambda \left(\frac{\sqrt{2}}{2} \sin\left(\frac{1}{\sqrt{2}}\right) - \cos\left(\frac{1}{\sqrt{2}}\right)\right) + \hat{f}(k)
\end{equation}

\begin{figure}[htb]
    \centering
    \includegraphics[width=0.5\linewidth]{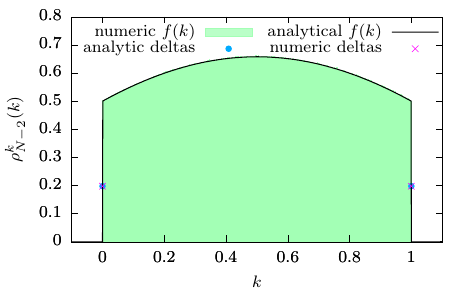} 
    \caption{\label{fig_num_test}Here we show the probability density function $\rho^{k}_{N-2}(k)$  over $k$ 
    to compare our analytical result from \autoref{derived_dist} to numeric measurements. The numeric 
    probability density was calculated by creating a histogram of 1000 bins and 
    then simulating for 200000000 time steps and counting each $k_{N-2}$ in the histogram. If $k=0$ or $k=1$ were reached exactly, we counted that for 
    the density of the deltas instead.}
\end{figure}

So all that is left is to determine the constants $\lambda$ and $\zeta$. 
Luckily we have two equations for that.
Firstly the constant part of \autoref{fk_equation} clearly needs to be 0, such that the equation is true ($\hat{f}(k) = \hat{f}(k)$).
Secondly, $\rho^{\hat{k}}_{N-2}(k)$ describes a probability density function and thus the integral
over it must be $1$.
\begin{align}
    \zeta &= -\lambda \left(\frac{\sqrt{2}}{2} \sin\left(\frac{1}{\sqrt{2}}\right) - \cos\left(\frac{1}{\sqrt{2}}\right)\right)\\
    1 &=\int_{-1/2}^{1/2} \rho^{\hat{k}}_{N-2}(k) dk = \sqrt{2} \lambda \sin \left(\frac{1}{\sqrt{2}}\right) + 2 \zeta
\end{align}
Solving this with the help of a computer algebra program gives:
\begin{align}
    \lambda &= \frac{1}{2} \sec\left(\frac{1}{\sqrt{2}}\right)\\
    \zeta &= \frac{1}{4} \left(2 - \sqrt{2} \tan \left(\frac{1}{\sqrt{2}}\right)\right)
\end{align}
Returning to our original coordinates, we have the following distribution for our stock,
which is valid for $k \in [0,1]$
\begin{equation}
    \label{derived_dist}
    \rho^{k}_{N-2}(k) = \left(\delta\left(k\right) + \delta(k-1)\right) \zeta + \lambda \cos\left(\sqrt{2} \left(k-\frac{1}{2}\right)\right) ~.
\end{equation}
Note that for $k \notin [0,1]$ the probability density is 0.
In \autoref{fig_num_test} we compare our analytic results with numerical measurements.
Clearly the numeric measurements match our analytic results exactly.

If we calculate the average stock value we get 
\begin{equation}
    \int_0^1 k \rho^{k}_{N-2}(k)  dk = \frac{1}{2}
\end{equation}
which is exactly what we expect for symmetry reasons.

Since we now know the exact probability distribution of $\rho^{k}_{N-2}(k)$ we can use
\autoref{I_n_2_needed_later_eq} to calculate the probability distribution $\rho^I_{N-2}(I)$.
The derivation itself is quite simple and can be done via the cumulative probability function.
It is, however, quite lengthy, so we omit the derivation here. The result for $0 \leq I \leq 1$ is:
\begin{align}
    \label{P_I_n_2_eq}
    \rho^I_{N-2}(I) &= (1-I)
    \left(\lambda \frac{\sin\left(\frac{2 I -1}{\sqrt{2}}\right) + \sin\left(\frac{1}{\sqrt{2}}\right)}{\sqrt{2}}
     + \zeta\right)\\
    \nonumber
    &+ \frac{1}{2} \lambda \left(\cos\left(\frac{2 I - 1}{\sqrt{2}}\right)- \cos\left(\frac{1}{\sqrt{2}}\right)\right)\\
    \nonumber
    &- \frac{\lambda (I-1) \sin\left(\frac{1}{\sqrt{2}}\right)}{\sqrt{2}} +\zeta - \zeta I + \frac{1}{2}
\end{align}

\begin{figure}[htb]
    \centering
    \includegraphics[width=0.5\linewidth]{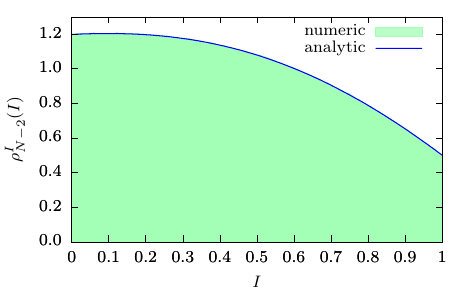} 
    \caption{\label{P_I_n_2_fit} Here we show the probability density function $\rho^I_{N-2}(I)$  over $I$ 
    to compare our analytical result from \autoref{P_I_n_2_eq} to numeric measurements. The numeric 
    probability density was calculated by creating a histogram of 1000 bins and 
    then simulating for 200000000 time steps and counting each $I_{N-2}$ in the histogram.
    }
\end{figure}

In \autoref{P_I_n_2_fit} we show the comparison of \autoref{P_I_n_2_eq} with the measured data.
Clearly this fits very well.

Next we can calculate the critical root demand $r^*$. For $N=1$ the network 
consists only of the leaf and we trivially have $r^*=\frac{1}{2}$.
For $s=1$ we can now also calculate the critical root demand for $N=2$:
\begin{equation}
    r^*=\int_0^1 I \rho^I_{N-2}(I)dI = \frac{7}{24} + \frac{\tan\left(\frac{1}{\sqrt{2}}\right)}{4 \sqrt{2}}
\end{equation}

\section{Numerical analysis for the open chain structure}

We measure the critical root demand $r^*$ for various $z_0$ and
different heights $h$, and use the fitting function defined in 
Eq.~12 in the main text. We fit the function to $N\geq 400$ for all $z_0$ except 
$z_0=1000$, for which we fitted the function to the data for $N\geq 5000$.
The coefficients of determination $R^2$ were all above 0.995,
indicating good fit quality, at least for large $N$.

\begin{figure}[htb]
    \centering
    \includegraphics[width=0.49\linewidth]{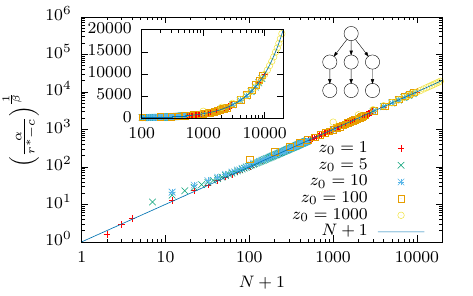} 
    \includegraphics[width=0.49\linewidth]{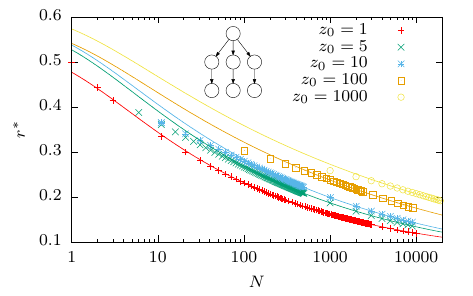}
    \caption{We measured $r^*$ over $N$ for $s=1$ and plot $\left(\frac{\alpha}{r^*-c}\right)^{\frac{1}{\beta}}$ 
    over $N+1$ (left plot), where 
    the fit parameters were obtained separately for each $z_0$ by fitting Eq.~12 (main text) to the measured $r^*$ (right plot).
    On the left plot we additionally show the identity function $N+1$ and the inset displays the same data but in half-logarithmic scale.
    The little graphs in the plots are a visual aid to indicate that these plots are about open chains.
    \label{DataCollapseMultiChain}
    }
\end{figure}

Rearranging Eq.~12 from the main text gives us 
$N(r^*)+1 = \left(\frac{\alpha}{r^*-c}\right)^{\frac{1}{\beta}}$,
which leads to a nice data collapse displayed in the left plot of
\autoref{DataCollapseMultiChain}.

\begin{figure}[htb]
    \centering
    \includegraphics[width=0.5\linewidth]{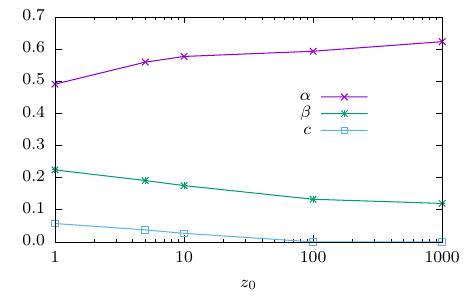} 
    \caption{\label{fit_params3} Fit parameters of \autoref{DataCollapseMultiChain}. The symbols represent the actual fit parameter whereas the lines 
    are just intended as visual aid.}
\end{figure}

The fits can be seen in \autoref{DataCollapseMultiChain} on the right.
Note that most fits 
were fitted to the data for $N\geq 400$, only for $z_0=1000$ did we fit to the data for $N\geq 5000$.
The fitting parameters can be found in \autoref{fit_params3}.

We observe that the larger $z_0$ is, the smaller the exponent $\beta$
becomes: adding a single node reduces the critical root demand rate less.
However, we also see that $c$ decreases with $z_0$, implying that networks with more 
chains converge to a lower critical demand rate $r^*$ for $N \to \infty$, which makes intuitive 
sense as there are more dependencies throughout the network.

\section{Line versus tree}

Next we compare the benefits of a tree structure consisting of a
root node and $N-1$ children with a line of $N$ nodes. For better comparability 
we use the same production capacity values in both cases, which can be achieved by using identical seeds 
in the simulation.

We use a root demand that is larger than 1 such that the system cannot satisfy this demand and thus 
the demand is never a limiting term in Eq.~(4).
We start with empty stocks and simulate for a generous $4 \times 10^7$
time steps to let the system equilibrate. 
Next we want to measure the difference in production for $4 \times 10^8$ time steps. 
More specifically we measure the sum of the production 
\begin{equation}
    I^{line}_{tot} = \sum_t I^{line} (t) \quad \text{and} \quad I^{tree}_{tot} = \sum_t I^{tree} (t) ~,
\end{equation}
which lets us measure the fraction of how much more the tree structure was able to produce compared to the line
\begin{equation}
    \chi = \frac{I^{tree}_{tot}  - I^{line}_{tot} }{I^{line}_{tot} }~.
\end{equation}

We measure this for different stock capacities $s$, using a different seed for each value of $s$ that we use.
The results can be found in the inset of Fig.~4 in the main text.

\begin{figure}[h!bt]
    \centering
    \includegraphics[width=0.5\linewidth]{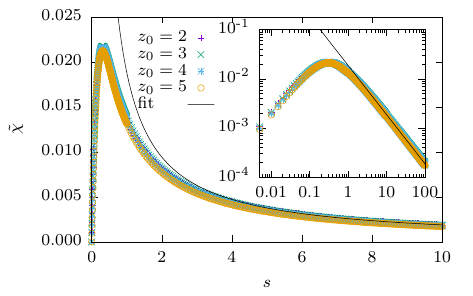} 
    \caption{\label{collapse_chi}We show $\tilde{\chi}$ over s for different values of $z_0$. 
    The inset displays the same data in logarithmic scale. We used $\beta=0.962$ to create the collapse with the help of \autoref{chi_collapse_eq}.
    We also fitted a power law to the data for $z_0=2$ for $s\geq 5$, which is displayed in the figure.
    }
\end{figure}

For $s=0$ we have $\chi=0$, which makes sense, because without the stocks the production of the root node only depends 
on the number of nodes as long as no node has multiple parents. 
For larger values of $s$ the relative difference $\chi$ increases at first until, for all measured trees, 
we reach a maximum at $s \approx 0.315$. Afterwards 
$\chi$ declines and approaches 0 for large $s$ (see also \autoref{collapse_chi}, which will be discussed below).
This makes sense, since for $s \to \infty$ our stocks are so large 
that their limit will never be exceeded, i.e., no product will ever be thrown away. 
Thus for all tree-like topologies the average production of the notes should approach the 
average production capacity, i.e., 
$\frac{1}{2}$, because we assumed that the demand is large enough 
to not be the limiting term in Eq.~(4). This is supported by \autoref{large_s_chain}
in the next section.

Now we want to see if we can estimate the scaling behavior. 
In the main text in Fig.~4
we already showed the relative production difference $\chi$ over $s$ for different $z_0$ in linear scale.
It looks like the shape of the functions are quite similar and they might only be rescaled.
Since the curve for $z_0=3$ seems to be roughly the curve of $z_0=2$ rescaled by 2,
whereas the curve for $z_0=4$ seems to be roughly the curve of $z_0=2$ rescaled by 3,
we try
\begin{equation}
    \label{chi_collapse_eq}
    \tilde{\chi}(s,z_0) = \frac{\chi(s,z_0)}{(z_0-1)^\beta}~.
\end{equation}
Overall this leads to a good collapse, as shown in \autoref{collapse_chi}.

For $s \geq 5$ we fitted a power law to the data of $z_0=2$, which gave us 
$\chi(s)=0.0188 s^{-1}$. The fit works well and gives us a coefficient of determination 
of $R^2 > 0.998$, indicating good fit quality.
The power law suggests that for $s \to \infty$ we have $\chi \to 0$, i.e.,
for large enough storage capacities $s$ both the line and the minimal tree have the same 
critical demand rate $r^*$.

\section{Additional figures \label{addFig}}

\begin{figure}[htb]
    \centering
    \includegraphics[width=0.5\linewidth]{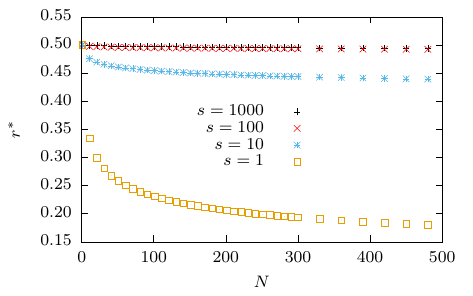} 
    \caption{\label{large_s_chain} We measured the critical demand rate $r^*$ over the number of nodes $N$ for different stock capacities $s$.
    The network topology was always a single chain.}
\end{figure}

In \autoref{large_s_chain} we show the measurement of $r^*$ over $N$ for large values of the maximal 
stock capacity $s$. As explained in the main text, we assume that we approach $r^*=\frac{1}{2}$ for $s\to \infty$.
The figure supports this hypothesis.

\begin{figure}[htb]
    \centering
    \includegraphics[width=0.5\linewidth]{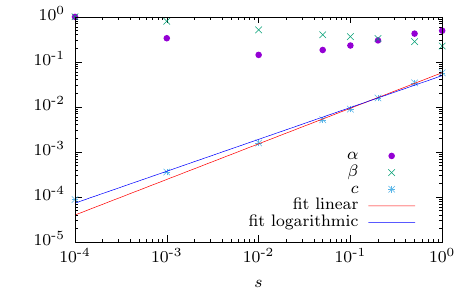} 
    \caption{\label{fit_params2} Fit parameters of Eq.~12 (main text) for Fig.~2 (main text). We also display two power law fits 
    for $c(s)$}
\end{figure}

Figure \autoref{fit_params2} shows the fit parameters from Fig.~2 from the main text. 
Since $c(s)$ looks like it could be a power law we fitted a power law through the data.
We did this twice, once by using weights that correspond to the logarithmic data,
for which we got $c(s)=0.049 s^{0.71}$, which gives us a coefficient of determination of $R^2=0.974$.
We also calculated a fit for the linear data and got $c(s)= 0.057, s^{0.79}$ 
and a coefficient of determination of $R^2>0.999$.

Overall we can see that for $s\to 0$ $\alpha$ and $\beta$ seem to converge to 1, whereas 
$c$ goes towards 0, as expected. This recovers the analytical result for $s=0$.

\begin{figure}[h!]
    \centering
    \includegraphics[width=0.5\linewidth]{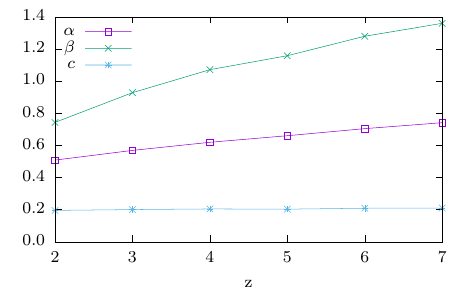} 
    \caption{\label{fit_params4} Fit parameters of the fits from Fig.~3 (main text). The symbols represent the actual fits, whereas the lines are 
    only meant as visual aid.}
\end{figure}

In \autoref{fit_params4} we show the fit parameters used to create Fig.~3 from the main text.
Note that in the limit of infinite height Eq.~13 (main text) reduces to $r^*(z,h=\infty)=c$.
Since $c$ seems to stay relatively constant, that would mean that for an infinite height $h$ the fits would 
predict a non-zero $r^*$ of about 0.2 for all measured $z$.

\bibliography{demand_paper.bib}